\documentclass[
reprint,
superscriptaddress,
amsmath,amssymb,
aps,
prl,
]{revtex4-1}

\usepackage{graphicx}
\usepackage{array}
\usepackage{dcolumn}
\usepackage{bm}
\usepackage{listings}
\usepackage{scalefnt}
\usepackage{xcolor}
\usepackage{booktabs}
\usepackage{multirow}
\usepackage{tensor}
\usepackage{hyperref}
\usepackage{mathtools}
\usepackage{textcomp}
\hypersetup{colorlinks=true, linkcolor={blue!80!black}, citecolor={blue!80!black}, urlcolor={blue!80!black}}
\usepackage{color}

\usepackage{physics}
\usepackage[pagewise]{lineno}

\begin{document}

\title{Observing the ``quantum Cheshire cat" effect with noninvasive weak measurement}

\author{Yosep Kim}
\affiliation{Department of Physics, Pohang University of Science and Technology (POSTECH), Pohang 37673, Korea}

\author{Dong-Gil Im}
\affiliation{Department of Physics, Pohang University of Science and Technology (POSTECH), Pohang 37673, Korea}

\author{Yong-Su Kim}
\affiliation{Center for Quantum Information, Korea Institute of Science and Technology (KIST), Seoul 02792, Korea}

\author{Sang-Wook Han}
\affiliation{Center for Quantum Information, Korea Institute of Science and Technology (KIST), Seoul 02792, Korea}

\author{Sung Moon}
\affiliation{Center for Quantum Information, Korea Institute of Science and Technology (KIST), Seoul 02792, Korea}

\author{Yoon-Ho Kim}
\email{yoonho72@gmail.com}
\affiliation{Department of Physics, Pohang University of Science and Technology (POSTECH), Pohang 37673, Korea}

\author{Young-Wook Cho}
\email{choyoungwook81@gmail.com}
\affiliation{Center for Quantum Information, Korea Institute of Science and Technology (KIST), Seoul 02792, Korea}

\date{\today}

\begin{abstract}
One of the common conceptions of nature, typically derived from the experiences with classical systems,  is that attributes of the matter coexist with the substance. In the quantum regime, however, the quantum particle itself and its physical property may be in spatial separation, known as the quantum Cheshire cat effect. While there have been several reports to date on the observation of the quantum Cheshire cat effect, all such experiments are based on  first-order interferometry and destructive projection measurement, thus allowing simple interpretation due to measurement-induced disturbance and also subject to trivial interpretation based on classical waves.   In this work, we report a genuine experimental observation of the quantum Cheshire cat effect with noninvasive weak quantum measurement as originally proposed.  The use of the weak-measurement probe  has allowed us to identify the location of the single-photon and that of the disembodied polarization state in a quantum interferometer. We furthermore elucidate the paradox of the quantum Cheshire cat effect as quantum interference of the transition amplitudes for the photon and the polarization state which are directly obtained from the measurement outcomes or the weak values.
\end{abstract}

\maketitle

Everyday experiences, typically derived from observing classical systems, shape up our common conceptions of nature. Quantum effects, on the other hand, often reveal peculiar counter-intuitive phenomena. Among such counter-intuitive phenomena, the quantum Cheshire cat effect, in which the quantum particle itself and its physical property are in spatial separation,  most vividly illustrates the stark difference between the quantum system and the classical system \cite{Aharonov13}. The disembodiment of the physical property (i.e., the state) from the particle itself is not only conceptually interesting, but  may also provide a novel way to avoid local decoherence on a certain physical state \cite{Richter18}. 

\begin{figure}[t]
\centering
\includegraphics[width=2.5 in]{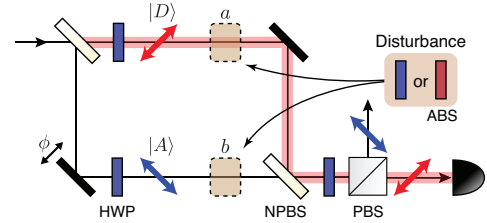}
\caption{Essential schematic of the quantum Cheshire cat experiments based on the Mach-Zehnder interferometer \cite{Denkmayr14,Ashby16}. A polarized beam is split into two paths with a non-polarizing beam splitter (NPBS). With half-wave plates (HWP), the polarizations in the upper and lower paths, respectively, are set at  $|D\rangle$  and $|A\rangle$. The polarizing beam splitter (PBS) ensures that only the $|D\rangle$ polarization reaches the detector. To probe the locations of the photon itself and its polarization state, an absorber (ABS) or a HWP is inserted in paths $a$ or $b$. 
}
\label{fig1}
\end{figure}

According to the original proposal for the quantum Cheshire cat effect \cite{Aharonov13}, the observation of the disembodiment effect requires probing the particle itself and the disembodied physical state with noninvasive weak quantum measurement  \cite{Banaszek01,Baek08}. Otherwise, i.e., if observed with projective measurement, measurement-induced disturbance plays a prominent role, thus making the quantum Cheshire cat effect argument inconsequential \cite{Cho19}. In the literature, the quantum Cheshire cat effect was reported via a neutron experiment \cite{Denkmayr14} and  a single-photon experiment \cite{Ashby16}, both based on essentially identical Mach-Zehnder interferometry setup shown in  Fig.~\ref{fig1}. An absorber (ABS) or a half-wave plate (HWP) inserted in one of the interferometric paths $a$ or $b$ acts as the probe for the particle itself or the disembodied physical state (i.e., spin or polarization). An argument in support of the quantum Cheshire cat effect may be made by observing the change of detection rate induced by the probe. For instance, if the photon travels along path $a$ (and its polarization state in path $b$),  an ABS inserted in path $b$ would not affect the detection probability at all.  
While these early experiments do provide some insights on the quantum Cheshire cat effect phenomenologically, they are based on simple first-order interferometry and destructive projection measurement \cite{Denkmayr14,Ashby16}. The original requirement of the noninvasive weak quantum measurement, necessary for unambiguous observation of the quantum Cheshire cat effect, was not satisfied \cite{Duprey18,Quach18,Sokolovski16,Correa15,Michielsen15}.  Therefore, the quantum Cheshire cat experiments reported to date are all prone to  interpretation based on measurement-induced disturbance and also subject to trivial interpretation based on classical waves. In fact, it is simple to reproduce the Mach-Zehnder type quantum Cheshire cat experiment by using classical light waves, thus requiring no quantum interpretation at all \cite{Stuckey16,Atherton15}. 


In this work, we report, to the best of our knowledge, the first  genuine experimental observation of the quantum Cheshire cat effect with noninvasive weak quantum measurement as originally proposed \cite{Aharonov13}. The location of the single-photon and that of the disembodied polarization state in a quantum interferometer have been identified by using the noninvasive weak-measurement probes. In our work, the weak measurement interaction is implemented by exploiting linear optical entangling gates via two-photon quantum interference, which cannot be explained classically \cite{Pryde05, Pryde04,Kiesel05}. We furthermore elucidate the paradox of the quantum Cheshire cat effect as quantum interference of the transition amplitudes for the photon and the polarization state which are directly obtained from the measurement outcomes or the weak values \cite{Sokolovski16}.

\begin{figure}[tp]
\centering
\includegraphics[width=3.2 in]{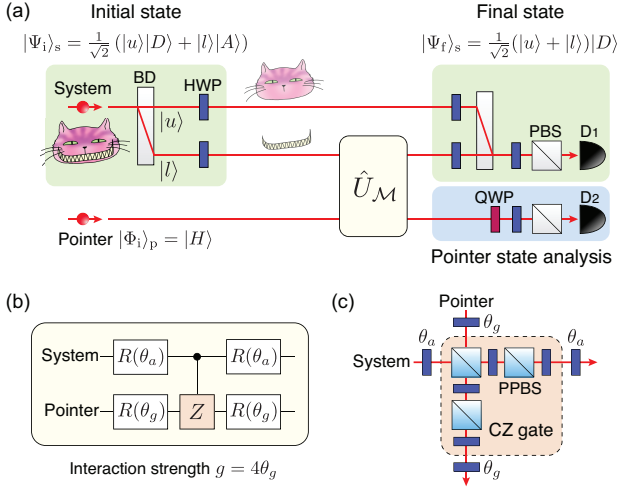}
\caption{(a) Experimental schematic to observe the quantum Cheshire cat effect. The system photon and its polarization are disembodied during the transit through the optical paths. The system states are prepared as presented in figure using HWP, beam displacers (BD), and a PBS. To probe the presence of the photon and the disembodied polarization state at the lower path, the system photon is weakly coupled to the pointer photon via the unitary interaction $\hat{U}_{\mathcal{M}}$. The quantum circuit for the unitary interaction  is shown in (b) and its quantum optical implementation is shown in (c). Note that the controlled-Z (CZ) gate is implemented via two-photon quantum interference at a partial polarizing beam splitter (PPBS) having polarization-dependent transmissions ($T_H=1$, $T_V=1/3$), and HWP for the CZ gate set at $45^{\circ}$. The pointer state is finally measured with  a quarter-wave plate (QWP), a HWP, and a PBS. 
 }
\label{fig2}
\end{figure}

We illustrate the experimental schematic to observe the quantum Cheshire cat effect in Fig.~\ref{fig2}(a). The system photon represents the Cheshire cat and its horizontal $|H\rangle$ and vertical $|V\rangle$ polarization states correspond to the status of her grin. The disembodiment of the grin (the polarization state) from the cat (the single photon) occurs as the system photon propagates between beam displacers (BD). The initial state of the photon in Fig.~\ref{fig2}(a) after the first BD and HWP is given as, \cite{Kim03}
\begin{equation}
|\Psi _{\mathrm{i}}\rangle_{\mathrm{s}} = \frac{1}{\sqrt{2}}( |u\rangle \otimes |D\rangle + |l\rangle \otimes |A\rangle),
\label{eq1}
\end{equation}
where $|D\rangle = (|H\rangle+|V\rangle)/\sqrt{2}$ and $|A\rangle = (|H\rangle-|V\rangle)/\sqrt{2}$. To probe the presence of the photon and its disembodied polarization state at the lower path, the pointer photon $|\Phi_{\mathrm{i}}\rangle_{\mathrm{p}}$, initialized in $|H\rangle$, is weakly coupled to the system photon for noninvasive measurement. Since there are only two possible optical paths, the information gained on the lower path $|l\rangle$ with the noninvasive measurement is sufficient to make complementary conclusions on the upper path  $|u\rangle$. 
The system photon is eventually subject to projection measurement in the basis $|\Psi_{\mathrm{f}}\rangle_{\mathrm{s}}$,
\begin{equation}
|\Psi_{\mathrm{f}}\rangle_{\mathrm{s}} =  \frac{1}{\sqrt{2}} (|u\rangle +  |l\rangle) \otimes  |D\rangle,
\label{eq2}
\end{equation}
and the state of the pointer photon is measured with  a QWP, a HWP, and a PBS. 

As shown in Fig.~\ref{fig2}(a), the system photon in the lower path $|l\rangle$ is weakly probed for the observable  $\hat{\Pi}_a=|a\rangle\langle a|$ via the measurement interaction of $\hat U_{\mathcal{M}}$~\cite{Pryde05, Pryde04}. The measurement interaction imparts a rotating operation $R(g/2)$ on the pointer state, conditioned on the system photon's polarization state $|a\rangle$, i.e., $\hat{U}_{\mathcal{M}}=(\hat{\mathbb{I}}-\hat{\Pi}_a)\otimes\hat{\mathbb{I}}+\hat{\Pi}_a\otimes R(g/2)$.  Here, $R(g/2)$ is defined by $R(g/2) |H\rangle \rightarrow \cos g |H\rangle + \sin g |V\rangle$ and $R(g/2) |V\rangle \rightarrow \sin g |H\rangle - \cos g |V\rangle$. The quantum circuit for the unitary interaction $\hat U_{\mathcal{M}}$ is shown in Fig.~\ref{fig2}(b) and its quantum optical implementation is shown in Fig.~\ref{fig2}(c).  The interaction strength $g=4\theta_g$ and the observable $\hat{\Pi}_a$ are set by the HWP angles $\theta_g$ and $\theta_a$. The controlled-Z (CZ) gate is implemented via two-photon quantum interference at a partial polarizing beam splitter (PPBS) having polarization-dependent transmissions ($T_H=1$, $T_V=1/3$), and HWP for the CZ gate set at $45^{\circ}$ \cite{Hong87,Kiesel05}. 

 The total unitary operation acting on the three-mode system-pointer state $|\Psi_{\mathrm{i}}\rangle_{\mathrm{s}} \otimes | \Phi_{\mathrm{i}}\rangle_{\mathrm{p}}$ is given by~\cite{Kim18}
\begin{eqnarray}
\hat{U}_{\mathrm{tot}}&&= \hat \Pi_u \otimes \hat {\mathbb{I}} \otimes \hat {\mathbb{I}}+\hat \Pi_l \otimes \hat U_{\mathcal{M}}\nonumber \\
&&= (\mathbb{\hat I} \otimes \mathbb{\hat I} - \hat \Pi_l \otimes \hat{\Pi}_a )\otimes \mathbb{\hat I} +  \hat \Pi_l \otimes \hat{\Pi}_a \otimes R(g/2).
\end{eqnarray}
For $g =\pi/2$, the projection operators on the system state, $\mathbb{\hat I} \otimes \mathbb{\hat I} -\hat \Pi_l \otimes \hat{\Pi}_a$ and $\hat \Pi_l \otimes \hat{\Pi}_a$, are perfectly distinguished by the pointer state and the outcomes of the projection measurement on the system can be extracted by analyzing the state of the pointer. In contrast, for $|g|\ll1$, the state of the system photon is weakly coupled to the pointer state, realizing the noninvasive weak measurement which is essential for the genuine observation of the quantum Cheshire cat effect.

In the limit of weak measurement, $|g| \ll 1$, the system-pointer evolution becomes $\hat{U}_{\mathrm{tot}}|\Psi_{\mathrm{i}}\rangle_{\mathrm{s}} | \Phi_{\mathrm{i}}\rangle_{\mathrm{p}}$ and is calculated to be,
\begin{equation}
\hat{U}_{\mathrm{tot}}|\Psi_{\mathrm{i}}\rangle_{\mathrm{s}}  | \Phi_{\mathrm{i}}\rangle_{\mathrm{p}}= |\Psi_{\mathrm{i}}\rangle_{\mathrm{s}} | H \rangle_{\mathrm{p}} + g  \hat \Pi_l \otimes \hat{\Pi}_a  |\Psi_{\mathrm{i}}\rangle_{\mathrm{s}}  | V\rangle_{\mathrm{p}}.\label{utot}
\end{equation}
Note that the state of the system photon is negligibly disturbed. Upon projection measurement of the system state onto the final state $|\Psi_{\mathrm{f}}\rangle_{\mathrm{s}}$, the pointer state becomes
\begin{equation}
|\Phi_{\mathrm{f}}\rangle_{\mathrm{p}} \propto |H\rangle_{\mathrm{p}} + g \langle \hat{\Pi}_l  \otimes \hat{\Pi}_a  \rangle_{\mathrm{w}} |V\rangle_{\mathrm{p}},
\label{pointer_f}
\end{equation}
where $\langle\hat{O}\rangle_{\mathrm{w}}$ indicates the weak value, defined as \cite{Aharonov88, Dressel14, Kim18,Piacentini161,Thekkadath16,Cho10}
\begin{equation}
\langle \hat O \rangle_{\mathrm{w}} = \frac{ \langle \Psi_{\mathrm{f}} | \hat O  |\Psi_{\mathrm{i}} \rangle  }   { \langle \Psi_{\mathrm{f}}   |\Psi_{\mathrm{i}} \rangle }\label{weakvalue}.
\end{equation}
The weak value is extracted by analyzing the final pointer state in Eq.~(\ref{pointer_f}) as follows,
\begin{eqnarray}
\langle \hat \sigma_{x}^a \rangle_{\mathrm{p}} &=& 2g\Re\langle \hat{\Pi}_l  \otimes \hat{\Pi}_a \rangle_{\mathrm{w}}, \label{eq7} \\
 \langle \hat \sigma_{y}^a \rangle_{\mathrm{p}} &=& 2g\Im\langle \hat{\Pi}_l  \otimes \hat{\Pi}_a \rangle_{\mathrm{w}} , \nonumber
\end{eqnarray}
where $\hat{\sigma}_{x}$ and $\hat{\sigma}_{y}$ are Pauli operators and the expectation values are defined as $ \langle \hat \sigma^a_k\rangle_{\mathrm{p}} = { {}_{\mathrm{p}}  \langle \Phi_{\mathrm{f}}| \hat \sigma_k | \Phi_{\mathrm{f}} \rangle_{\mathrm{p}}} /  {}_{\mathrm{p}} \langle \Phi_{\mathrm{f}}| \Phi_{\mathrm{f}} \rangle_{\mathrm{p}}$.

In the experiment, the system and the pointer photons at 780 nm are produced via spontaneous parametric down conversion from a type-II beta-barium borate (BBO) crystal pumped by a 390 nm pulsed laser. The single-photons are delivered to the experimental setup shown in Fig.~\ref{fig2}(a) via the single-mode optical fibers and, as described earlier, the measurement interaction $\hat{U}_\mathcal{M}$ is based on two-photon quantum interference. To ensure high degree of spectral indistinguishability, necessary for high-visibility quantum interference, between the two single-photons, 1-nm bandwidth interference filters are placed in front of the detector D$_1$ and D$_2$.


For the quantum Cheshire cat effect, the relevant observables are $\hat \Pi_l \otimes \mathbb{\hat I} $ and $\hat \Pi_l \otimes \hat \sigma_z$, which represent the existence of the system photon itself and the presence of the photon's polarization state, respectively, in the lower path $|l\rangle$. The measurement outcomes $\langle \hat \Pi_l \otimes \mathbb{\hat I} \rangle_{\mathrm{w}}= 0$ and $ \langle \hat \Pi_l \otimes \hat \sigma_z \rangle_{\mathrm{w}}=1$ indicates the observation of the quantum Cheshire cat effect: the polarization state is found in the path in which the system photon does not exist. We obtain the weak values of $\hat \Pi_l \otimes \mathbb{\hat I}$ and $\hat \Pi_l \otimes \hat \sigma_z$ from the linear combinations of the weak values of $\hat \Pi_l \otimes \hat \Pi_H$ and $ \hat \Pi_l \otimes \hat \Pi_V$ by making use of the relations $\mathbb{\hat I}= \hat{\Pi}_H + \hat{\Pi}_V$ and $\hat \sigma_z = \hat{\Pi}_H- \hat{\Pi}_V$ as follows:
\begin{eqnarray}
\langle\hat{\Pi}_l\otimes\hat{\mathbb{I}}\rangle_{\mathrm{w}}=\langle\hat{\Pi}_l\otimes\hat{\Pi}_H\rangle_{\mathrm{w}}+\langle\hat{\Pi}_l\otimes\hat{\Pi}_V\rangle_{\mathrm{w}}, \\
\langle\hat{\Pi}_l\otimes\hat{\sigma}_z\rangle_{\mathrm{w}}=\langle\hat{\Pi}_l\otimes\hat{\Pi}_H\rangle_{\mathrm{w}}-\langle\hat{\Pi}_l\otimes\hat{\Pi}_V\rangle_{\mathrm{w}}. \nonumber
 \end{eqnarray} 
According to Eq.~(\ref{eq7}), the real and imaginary parts of $\langle \hat{\Pi}_l \otimes \hat{\mathbb{I}}\rangle_{\mathrm{w}}$ are obtained from $(\langle\hat{\sigma}_x^H\rangle+\langle\hat{\sigma}_x^V\rangle)/2g$ and $(\langle\hat{\sigma}_y^H\rangle+\langle\hat{\sigma}_y^V\rangle)/2g$ at $|g|\ll1$, and $\langle \hat{\Pi}_l \otimes \hat{\sigma}_z\rangle_{\mathrm{w}}$ is estimated similarly.

\begin{figure}[tp]
\centering
\includegraphics[width=3.2 in]{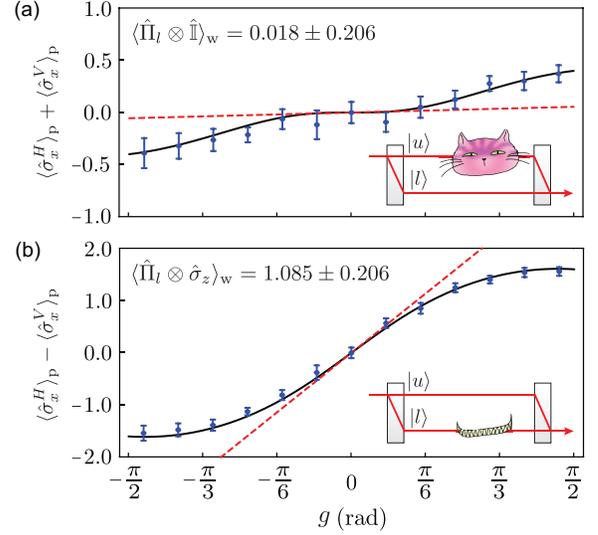}
\caption{Experimental observation of the quantum Cheshire cat effect. The observation is realized by measuring weak values. Measurements of pointer states (solid circle) are recorded as a function of $g$ for observables: (a) $\hat \Pi_l \otimes \mathbb{\hat I} $ and (b) $\hat \Pi_l \otimes \hat \sigma_x $. Note that the imaginary parts have zero value, so the results for $\langle \hat{\sigma}_y^a \rangle_{\mathrm{p}}$ are not presented. One standard deviation due to Poissonian counting statistics are considered as error bars. The black solid lines are the exact theoretical predictions. For given observables, weak values are extracted by taking the first order derivative with polynomial curve fit at $g$=0. The measurement results of $\langle \hat \Pi_l \otimes \mathbb{\hat I} \rangle_{\mathrm{w}} =0.018\pm0.206 $ and $\langle \hat \Pi_l \otimes \hat \sigma_z \rangle_{\mathrm{w}} =1.085\pm0.206$ indicate the quantum Cheshire cat effect that the physical property (polarization) can be found in the path where the physical carrier (photon) does not exist.
}\label{fig3}
\end{figure}

The experimental confirmation for the observation of the quantum Cheshire cat effect is shown in Fig.~\ref{fig3} in which the pointer measurements $\langle \hat \sigma_x^H \rangle_{\mathrm{p}}+\langle \hat \sigma_x^V \rangle_{\mathrm{p}}$ and $\langle \hat \sigma_x^H \rangle_{\mathrm{p}}-\langle \hat \sigma_x^V \rangle_{\mathrm{p}}$ are shown as a function of the measurement strength $g$. Each measurement interaction for $\langle \hat \sigma_x^H \rangle_{\mathrm{p}}$ and $\langle \hat \sigma_x^V \rangle_{\mathrm{p}}$ is implemented by setting the HWP angle $\theta_a$ in Fig.~\ref{fig2}(c) as 45$^{\circ}$ and 0$^{\circ}$, respectively. Then, the pointer state, conditioned on the projection measurement of the system onto the state $|\Psi_{\mathrm{f}}\rangle_{\mathrm{s}}$ at detector D$_1$, is analyzed from the coincident detection events of D$_1$ and D$_2$ with the set of a QWP, a HWP, and a PBS at detector D$_2$. The expectation values of $\langle \hat \sigma_x^H \rangle_{\mathrm{p}}$ and $\langle \hat \sigma_x^V \rangle_{\mathrm{p}}$ are obtained at each $g$, and the sum and the difference are given as the data points in Figs.~\ref{fig3}(a) and (b) for the real parts of $\langle\hat \Pi_l \otimes \mathbb{\hat I}\rangle_{\mathrm{w}}$ and $ \langle\hat \Pi_l \otimes \hat \sigma_z\rangle_{\mathrm{w}}$, respectively. Note that the imaginary parts have zero value, so the results for $\langle \hat{\sigma}_y^a \rangle_{\mathrm{p}}$ are not presented.

The weak values are extracted from the experimental data in Fig.~\ref{fig3} from the slope at $g=0$ by using the polynomial fit to the data according to the relation in Eq.~(\ref{eq7}). The experimentally obtained weak values are  $\langle \hat \Pi_l \otimes \mathbb{\hat I} \rangle_{\mathrm{w}} =0.018\pm0.206 $ and $\langle \hat \Pi_l \otimes \hat \sigma_z \rangle_{\mathrm{w}} =1.085\pm0.206$ are in good agreement with the theoretical prediction and clearly demonstrate the quantum Cheshire cat effect.

\begin{figure}[tp]
\centering
\includegraphics[width=2.9 in]{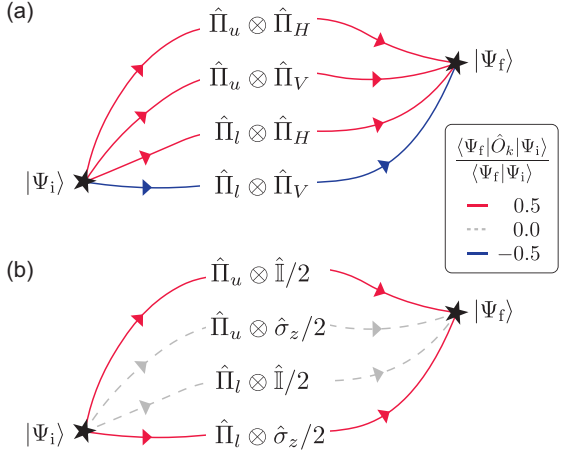}
\caption{Conceptual Feynman diagrams for the transition from $|\Psi_{\mathrm{i}}\rangle$ to $|\Psi_{\mathrm{f}}\rangle$. The virtual paths defined by $\hat{O}_k$ can be arbitrarily set to satisfy $\sum_{k}\hat{O}_k=\hat{\mathbb{I}}$ so that $\langle\Psi_{\mathrm{f}}|\Psi_{\mathrm{i}}\rangle=\sum_k\langle\Psi_{\mathrm{f}}|\hat{O}_k|\Psi_{\mathrm{i}}\rangle$. (a) $\hat{O}_k=\{\hat{\Pi}_u\otimes\hat{\Pi}_H, \hat{\Pi}_u\otimes\hat{\Pi}_V, \hat{\Pi}_l\otimes\hat{\Pi}_H, \hat{\Pi}_l\otimes\hat{\Pi}_V\}$, (b) $\hat{O}_k=\{\hat{\Pi}_u\otimes\hat{\mathbb{I}}/2, \hat{\Pi}_u\otimes\hat{\sigma}_z/2, \hat{\Pi}_l\otimes\hat{\mathbb{I}}/2, \hat{\Pi}_l\otimes\hat{\sigma}_z/2\}$, where $\hat{\Pi}_u$ and $\hat{\Pi}_l$ represent the spatial modes of the state and $\hat{\Pi}_H$ and $\hat{\Pi}_V$ represent the polarization modes of the state. The transition amplitudes along the virtual paths in (a) and (b) are related with each other due to $\hat{\mathbb{I}}=\hat{\Pi}_H+\hat{\Pi}_V$ and $\hat{\sigma}_z=\hat{\Pi}_H-\hat{\Pi}_V$. The line color shows the normalized transition amplitude of each virtual paths by the total transition amplitude $\langle\Psi_{\mathrm{f}}|\Psi_{\mathrm{i}}\rangle$ for the initial and final states in Eqs.~(\ref{eq1}) and~(\ref{eq2}).}
\label{fig4}
\end{figure}

While the quantum Cheshire cat effect may look paradoxical, we may interpret the effect as quantum interference of the transition amplitudes for the photon and the polarization state. The weak value of Eq.~(\ref{weakvalue}), formally, can be interpreted as the transition amplitude $\langle \Psi_{\mathrm{f}} | \hat O  |\Psi_{\mathrm{i}} \rangle$ along the virtual path defined by $\hat O$ from the initial state $|\Psi_{\mathrm{i}}\rangle$ to the final state $|\Psi_{\mathrm{f}}\rangle$, which is normalized by the total transition amplitude $\langle\Psi_{\mathrm{f}}|\Psi_{\mathrm{i}}\rangle$~\cite{Sokolovski16}. Considering the spatial modes $\hat{\Pi}_u$ and $\hat{\Pi}_l$ and the polarization modes $\hat{\Pi}_H$ and $\hat{\Pi}_V$, there are four possible virtual transition paths as follows, see Fig.~\ref{fig4}(a).  
\begin{equation}
\hat \Pi_u \otimes \hat{\Pi}_H,\ \hat \Pi_u \otimes \hat{\Pi}_V,\ \hat \Pi_l \otimes \hat{\Pi}_H,\ \hat \Pi_l \otimes  \hat{\Pi}_V.
\label{set1}
\end{equation}
The weak values, namely the normalized transition amplitudes $\langle\hat{O}\rangle_{\textrm{w}}=\langle \Psi_{\mathrm{f}} | \hat O  |\Psi_{\mathrm{i}} \rangle/\langle \Psi_{\mathrm{f}} |\Psi_{\mathrm{i}} \rangle$, for the initial and final states in Eqs.~(\ref{eq1}) and~(\ref{eq2}) are given as,
\begin{eqnarray}
&&\langle \hat \Pi_u \otimes \hat{\Pi}_H \rangle_{\mathrm{w}}= 0.5,~~ \langle \hat \Pi_u \otimes  \hat{\Pi}_V \rangle_{\mathrm{w}}=0.5, \nonumber \\
&&\langle \hat \Pi_l \otimes \hat{\Pi}_H\rangle_{\mathrm{w}}= 0.5,~~ \langle \hat \Pi_l \otimes \hat{\Pi}_V \rangle_{\mathrm{w}}=-0.5.
\label{amp_set1}
\end{eqnarray}
Note that the sum of all the normalized transition amplitudes is equal to unity because the observables sum to the identity operator.

Similarly, as shown in Fig.~\ref{fig4}(b), another complete set of virtual transition paths exists as the following observables, 
\begin{eqnarray}
\hat \Pi_u \otimes \mathbb{\hat I}/2,~~\hat \Pi_l \otimes \mathbb{\hat I}/2,~~\hat \Pi_u \otimes \hat \sigma_z/2,~~\hat \Pi_l \otimes \hat \sigma_z/2,
\label{set2}
\end{eqnarray}
where the observables also sum to the identity operator.
The corresponding normalized transition amplitudes for the initial and final states in Eqs.~(\ref{eq1}) and~(\ref{eq2}) are calculated as,
\begin{eqnarray}
&&\langle \hat \Pi_u \otimes \mathbb{\hat I}/2 \rangle_{\mathrm{w}}= 0.5,~~ \langle \hat \Pi_u \otimes \hat \sigma_z/2 \rangle_{\mathrm{w}}=0, \nonumber \\
&&\langle \hat \Pi_l \otimes \mathbb{\hat I}/2 \rangle_{\mathrm{w}}= 0,~~ \langle \hat \Pi_l \otimes \hat \sigma_z/2 \rangle_{\mathrm{w}}=0.5.
\label{amp_set2}
\end{eqnarray}
The normalized transition amplitudes signify that the system photon can be found in the only upper path $|u\rangle$ while the polarization appears in the only lower path $|l\rangle$ during the transition. This paradoxical results can be understood as the interference between the fundamental transition amplitudes in Eq.~(\ref{amp_set1})~\cite{Sokolovski16}. The observables in Eq.~(\ref{set2}) can be expressed as the linear combination of the observables in Eq.~(\ref{set1}), e.g., $\hat{\Pi}_u\otimes\mathbb{\hat I}= \hat{\Pi}_u\otimes\hat{\Pi}_H + \hat{\Pi}_u\otimes\hat{\Pi}_V$ and $\hat{\Pi}_u\otimes \hat \sigma_z= \hat{\Pi}_u\otimes\hat{\Pi}_H - \hat{\Pi}_u\otimes\hat{\Pi}_V$. Consequently, it is possible to interpret the transition amplitudes in Eq.~(\ref{amp_set2}) as the outcomes of constructive and destructive interference between the transition amplitudes in Eq.~(\ref{amp_set1}). For instance, the null transition amplitude for the observable $\hat \Pi_l \otimes \mathbb{\hat I}$ is because of a destructive interference of two transition paths $\hat{\Pi}_l\otimes\hat{\Pi}_H$ and $\hat{\Pi}_l\otimes\hat{\Pi}_V$. 

While the weak quantum measurement has allowed us to investigate the counterfactual quantum Cheshire cat effect, it is important to remember that, unlike projection measurement, irreversible quantum state collapse has not occurred. If one measures the transition probability via strong projection measurement, not the amplitudes of the observables in Eq.~(\ref{set1}), the post-measurement state would be fully collapsed into one of the eigenstates of the measurement observable and, therefore, the sum of the transition probabilities $|\langle\Psi_{\mathrm{f}}|\hat{\Pi}_l\otimes\hat{\Pi}_H|\Psi_{\mathrm{i}}\rangle|^2+|\langle\Psi_{\mathrm{f}}|\hat{\Pi}_l\otimes\hat{\Pi}_V|\Psi_{\mathrm{i}}\rangle|^2$ would not exhibit quantum interference. This outcome  is also completely consistent with the classical wave theory.

To conclude, we have reported, to the best of our knowledge, the first genuine experimental observation of the quantum Cheshire cat effect. As suggested in the original proposal \cite{Aharonov13}, we have probed the photon's existence and its polarization property using a noninvasive weak measuring apparatus during the state transition. The noninvasive weak measuring apparatus was realized by coupling the quantum system with the quantum pointer by utilizing another single-photon. The quantum pointer reveals the quantum Cheshire cat effect in the framework of the weak value measurement and the apparent quantum Cheshire cat paradox was explained as quantum interference of virtual transition paths. Our experimental apparatus can be applied to investigate other paradoxical phenomena based on weak value such as Hardy's paradox \cite{Aharonov02,Lundeen09, Yokota09}, Leggett-Garg inequality \cite{Leggett85,Palacios10,Goggin11}, and quantum pigeonhole paradox \cite{Aharonov16,Chen19}.

\vspace{2mm}

This work was supported by the National Research Foundation of Korea (Grant Nos. 2019R1A2C3004812, 2019M3E4A107866011, and 2019M3E4A1079777) and KIST institutional programs (Project No. 2E30620). Y.K. acknowledges support from the Global Ph. D. Fellowship by the National Research Foundation of Korea (Grant No. 2015H1A2A1033028).




\end{document}